# Resilient Service Embedding
# In IoT Networks

Haider Qays Al-Shammari, Ahmed Lawey, Taisir El-Gorashi and Jaafar M. H. Elmirghani

*Abstract*— The Internet of Things (IoT) has been applied to a large number of heterogeneous devices and is used in the deployment of a variety of applications on the basis of its distributed open architecture. The majority of these IoT devices are battery-powered and are interconnected via a wireless network. IoT devices may be used to carry out critical tasks. Thus, the IoT network requires a resilient architecture that supports semantic search, failure discovery, data recovery, and dynamic and autonomous network maintenance.

In this paper, we present a new resilience scheme for IoT networks. We evaluate the proposed scheme in terms of its power consumption and data delivery time, and then compare the results with those of recent resilience schemes such as schemes based on redundancy and replication. The proposed framework was optimized using mixed integer linear programming and real-time heuristics were developed, thus embedding a virtual layer into a physical layer based on a service-oriented architecture (SOA). The proposed framework offers different combinations of packet resilience in terms of recovering the lost data by using end-to-end mechanisms. We further analyzed these schemes by investigating the power consumption, data delivery time, and network overhead of these techniques. The results showed that the proposed splitting technique enhanced the network performance by reducing the power consumption and the data delivery time of service embedding by selecting energy-efficient nodes and routes in IoT networks.

**Keywords: IoT, SOA, Resilience, Energy Efficiency, Traffic Latency, Queuing, MILP, Smart Buildings.**

## I. INTRODUCTION

The Internet of Things (IoT) is an emerging technology that can support different devices connected to the Internet to service ubiquitous and pervasive applications. The IoT facilitates the connection and interaction between smart objects and their services and the interconnection of embedded devices (e.g. sensors and actuators) using the Internet infrastructure. IoT can also enable a range of services/applications offered to smart buildings [1]. In a smart building paradigm, the embedded sensors collect data from certain specific places and send them to the controller for processing and to make decisions seamlessly and efficiently. The collected data has to be sent to the cloud, fog, or data centre, as these devices carry out actions defined by the IoT services. The combination of a smart building and IoT has been used in several paradigms and research studies [2] and poses several challenges when the reliability of services has to be guaranteed. Some of the key challenges are due to the vulnerabilities of the interconnectivity and the interdependencies of the devices and applications. Physical connectivity and hardware limitations can lead to unexpected system failures caused by failures in the interconnected the networks. In addition to the network connectivity, the large heterogeneity of network access technologies increases the complexity of the network and can cause deployment problems in the communication domain. The political and social acceptance challenges may appear as another type of challenge in the form of privacy and civil rights concerns because of the right to access and use the information in the smart building. Furthermore, economic challenges can constrain the financial budget for the replacement and deployment of the new technologies [3].

The majority of IoT devices have wireless connectivity, and thus, survivability and failure tolerance are important considerations. As the IoT plays a significant role in smart building projects, the traffic resilience of an IoT network is also considered an important factor in the design of smart building projects [4, 5]. This resilience is a significant consideration in various engineering, scientific, and social applications, as it has a considerable magnitude in ultra-large-scale systems [6]. Theoretically, there are many definitions of resilience; it can be defined as the capability of a system to accomplish its operation in an appropriate manner notwithstanding disruptions and regain its performance after a temporary system failure. In communication systems, the adverse disruption is a prospective consideration, and these systems are expected to operate even under adverse disruptions and to rapidly recover to their full functional services [7].

The IoT concept promises to support a large number of services ranging from those in smart homes to the automation of industries and public utilities. However, the increase in the number of these deployments has posed a significant challenge relating to the design of a resilient manner IoT architecture. The IoT nodes are prone to unexpected failures and malicious attacks, i.e. various types of damage, unreliable wireless connections, limited transmission power, computing ability, and storage space. The IoT paradigm consists of a heterogeneous combination of Internet-connected devices. In addition, traffic routing in IoT networks mainly relies on routing protocols for low-power and lossy networks (RPL). RPL protocols are designed to find a single route between the source and the destination nodes [8]. They can thus affect the services delivered by the networks, due, for example, to intermittent node faults, dropped radio links or a change in the network connectivity in addition to their vulnerability to attacks [9].

In general, there are many definitions of 'resilience'. The most common one is that it is the ability to operate and maintain a process with an acceptable level of service when facing various faults [3]. The work in [1] defined network resilience as the ability to have at least one operational backup path within a certain minimum time interval when at least one node on the primary path fails. Practically, traffic resilience can be measured by the time required by the network to resume its normal operation after being subjected to disruption [3, 10]. Consequently, it is complicated to estimate network resilience in terms of the quantitative value of the network resilience. Another key aspect is the number of failed nodes or links that the network can endure while maintaining its performance [8].

In IoT networks, traffic routing mainly relies on RPL, engineered by IETF in 2009 [11]. The RPL protocol is considered to be the de facto routing protocol for the IoT, because of its fit to the IoT requirements and it contributions to the improvement of the communications with other standards in order to provide a baseline architecture for IoT. RPL was designed to find a single route between the source and the destination nodes. Therefore, network resilience is important in this context. Its goal should be to improve the network's ability to handle faults and restore its operation, and does not necessarily imply that the system is very difficult to degrade [10]. Resource constraints, energy limitations, unreliability of wireless links, and single-path routing technique are factors that degrade the IoT network resilience and performance. In order to overcome these factors, many research groups have proposed multi-path solutions for the routing protocol in IoT networks.

Among these traffic routing protocols, a popular resilient technique for link failure recovery is multipath routing, where a set of multiple paths between the source and the destination are selected to ensure traffic delivery. This technique has the advantage of high resilience but with varying energy consumption and link capacity.

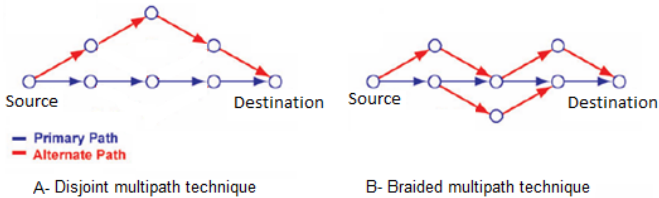

Fig. 1: Multipath techniques

Multipath methods have two main techniques to create their multipath network, as shown in Fig. 1. The first is disjoint multipath. In this technique, a number of paths with independent nodes/links are created as alternatives to the primary path, and thus, a failure in any or all the nodes/links on the primary path does not affect any of the alternative paths. The second technique is braided multipath. In this technique, the alternative paths partially overlay the primary path, as shown in Fig. 1-B, and thus, if any node on the primary path fails, new path discovery is required, which introduces an additional overhead [12, 13]. Resilient routing protocols in IoT networks [3], [22] are categorized into three types on the basis of the path finding methodology. The first method is called proactive routing, where all the paths are selected beforehand in the routing table, and the second method is reactive routing, where all the paths are selected on demand and updated in the routing table. The third method is hybrid routing and depends on both previous methods [13]. This leads to a probabilistic approach that assumes that the network can tolerate at most $n$ failed nodes where $0 < n < k$, in a $k$-connected network. The term $k$-connected network denotes the fact that the network can preserve its node connectivity after removing no more than k − 1 nodes [14-16]. The value of $k$ is an indicator of the network resilience, where a high value of $k$ denotes high network resilience.

In this study, we investigated various resilience schemes for IoT nodes and for the traffic they generate. We evaluated the performance and the implications of these schemes, such as the data delivery time and the energy consumption. We defined Business Processes (BP) as a collection of functions provided by IoT nodes and links such as data processing, data storage, sensing, actuation and communication. An IoT service to be embedded in an IoT network may contain a set of BPs interconnected in a given topology. We formulated the problem of finding the optimal set of IoT nodes and links to embed BPs into the IoT layer as an optimization problem, with the objective of minimizing both the total power consumption and the traffic latency. This problem was formulated using mixed integer linear programming (MILP). We benefit from our track record in energy efficiency and networks virtualization, and track record in IoT service embedding [17], [18], [19].

The authors of [20], [21] developed schemes to enhance the energy efficiency of IoT networks, while the authors of [22], [23] considered the virtualization of such networks. Processing the sensor data and the use of analytics based on such big data was surveyed in [24], while the author in [25] considered these analytics for effective actuation in the network. Greening these big data networks was introduced and discussed in [26], [27] whereas improving the energy efficiency of the clouds and their interconnecting networks that process the IoT data was evaluated in [28], [29] with the energy efficiency of content sharing optimized in [29] and[30]. The energy efficiency of the networks supporting different services was optimized in [31-39]. Resilience is essential for a range of services, hence [40] and [41] introduce strategies to improve resilience with energy efficiency. The work in [42] considers the use of big data analytics based on data collected from IoT networks to improve the quality of service offered to users, while [43, 44] consider ways to embed functions in the network while maximising energy efficiency.

In our previous work, [17], [18], [19], we have evaluated the energy efficiency of IoT service embedding with QoS parameters including traffic queuing latency, while in this work, we introduce resilience for the first time to IoT service embedding and evaluate the energy efficiency with level of resilience considering a range of scenarios.

The rest of this paper is organized as follows: In Section II, we review traffic resilience in IoT networks. In Section III, we propose our resilience framework, compare it with recent techniques, and introduce our new technique for the evaluation of resilience of service embedding. In Section IV, we discuss the results obtained. Finally, Section V concludes this paper.

## II. RESILIENT SERVICE EMBEDDING IN IOT NETWORK

We developed a framework that enhances the resilience of service embedding in IoT networks (for example in a smart building setting). We therefore introduce resilience to service embedding, where we proposed and studied IoT service embedding in work [17], [18], [19]. This framework aims to structure a network such that it has an acceptable level of fault tolerance and introduces the ability to restore from a node or link failure in the network. The framework proposes multilevel resilience schemes, where each probable type of failure (i.e. sensor, controller, or link failure) requires an appropriate level of failure recovery. We evaluated the proposed resilience levels by considering their impact on the end-to-end service delay and the energy consumption. The proposed resilience levels are as follows:

### A. Resilient service embedding with node coexistence constraint.

We considered service embedding with a coexistence constraint as the basic level of resilience. This scheme is considered to be the basic solution for a network with a probable temporary failure, i.e. data collision or packet drop.

This resilience scheme is based on a single path between the source and the destination nodes, where the source node ensures the recovery of lost packets by retransmitting them until an acknowledgement is received from the destination node. This scheme has the disadvantages of additional transmission overheads and high network congestion.

### B. Resilient service embedding with sensor–actuator node redundancy.

To enhance the resilience of IoT networks, we introduce redundant nodes and links for the sensor and actuator nodes. This redundancy scheme enhances the infrastructure's resilience against a service failure or disruptive attacks. We considered the redundant sensing and actuating nodes for accuracy and data fidelity in addition to the resilience concern.

### C. Resilient service embedding with all-node redundancy.

In many services, resilience has significant importance, such as fire protection and security services in public buildings. As the cost of the service components (e.g. nodes and energy consumption) is a non-substantial concern, a new feasible scheme based on the allocation of redundant components for all the nodes enables end-to-end traffic routing with multiple paths capability.

### D. Resilient service embedding with traffic redundancy.

This scheme is related to traffic resilience and is based on setting up multiple paths between the source and the destination nodes. One of these paths is considered the main or primary path to route the traffic between the nodes, while one or more other paths are considered the alternative or backup paths. These paths are used to recover from a traffic failure of the primary path and are sustained by sending a 'Keep-alive' signal continuously over the paths. When a primary path has a failure, the intermediate node sends back the data packet to the source node and sends a failure report to the destination node. As a result, the source and the destination nodes remove the failed path information from the routing table and switch the traffic to an alternative path.

### E. Resilient service embedding with traffic replication.

This scheme fulfils the requirement of resilient traffic by sending multiple replicas of the data over selected multiple paths from the source node to the destination node. This technique has the advantages of high packet delivery ratio with low data delivery time, and there is no need for signaling for state maintenance between the source node and the destination node, because even in the case of a partial data packet loss, the destination node can recover the packet from the other copies of the packet. Replication achieves high resilience but at the cost of high energy consumption that arises because of the added traffic and traffic overheads at each node along the network.

### F. Resilient service embedding with traffic splitting.

Here, we propose a technique where traffic is split from the source node to the destination node in two paths, where each path routes 50% of the data traffic, and the 'Keep-alive' signal is redirected on the same path. When a failure is encountered on one path, the source resends the undelivered data, which does not exceed 50% of the original data, of the failed path on the second path. Consequently, this scheme saves both energy and delivery time. We propose the use of a braided multipath technique in our framework. In this technique, the alternative nodes partially overlay the nodes of the primary path to avoid service blockage.

## III. MILP FRAMEWORK OF RESILIENT-ENERGY EFFICIENT SERVICE EMBEDDING IN IoT NETWORKS

In this section, we introduce our framework developed to embed services in IoT networks in a smart building setting. This framework is based on a MILP optimization model with the objective of minimizing the total power consumption and the traffic mean latency of the service embedding in IoT networks and enhancing the node/traffic resilience level.

### A. Framework definitions

Before introducing the framework, we define the following sets, parameters, and variables:

| Sets | |
|------|--|
| $B$ | Set of business processes (BPs) in the virtual layer |
| $V$ | Set of virtual nodes in each BP |
| $VN_{ia}$ | Set of neighbors of each virtual node in each BP ($i \in B, a \in V$) |
| P | Set of IoT nodes in the physical layer |
| $PN_c$ | Set of neighbors of IoT nodes ($c \in P$) |
| F | Set of functions supported by IoT nodes |
| $Z$ | Set of zones in the IoT physical layer |
| $\lambda$ | Set of arrival rates |
| $W_j$ | Set of mean latency per arrival rate ($j \in \lambda$) in ms per packet |
| **Parameters** | |
| $V_{ian}^{FUNC}$ | $V_{ian}^{FUNC} = 1$ If virtual node $a$ in BP $i$ requires the function $n$, $V_{ian}^{FUNC} = 0$ otherwise |
| $V_{iaz}^{ZONE}$ | $V_{iaz}^{ZONE} = 1$ If virtual node $a$ in BP $i$ requires zone $z$, $V_{iaz}^{ZONE} = 0$ otherwise |
| $V_{ia}^{MCU}$ | Processing requirement of the virtual node $a$ in BP $i$ in MHz |
| $V_{ia}^{RAM}$ | Memory requirement of the virtual node $a$ in BP $i$ in kB |
| $V_{iab}^{TRFH}$ | Traffic demand between the virtual node pair $(a, b)$ in $BP$ $i$ in kb/s |
| $P_{cn}^{FUNC}$ | $P_{cn}^{FUNC} = 1$ If IoT node $c$ can provide the function $n$, $P_{cn}^{FUNC} = 0$ otherwise. |
| $P_{cz}^{ZONE}$ | $P_{cz}^{ZONE} = 1$ If the IoT node $c$ is located in zone $z$, $P_{cz}^{ZONE} = 0$ otherwise. |
| $P_c^{MCU}$ | Processing capability of the IoT node $c$ in MHz. |
| $P_c^{RAM}$ | Memory capability of the IoT node $c$ in kB. |
| $P_{ef}^{DIST}$ | Distance between the neighboring IoT nodes pair $(e, f)$ in meters. |
| $P_c^{IDLE}$ | Idle processor power in each IoT node $c$ in mW. |
| $P_c^{MAXG}$ | Maximum processor power consumption in each IoT node $c$ in mW. |
| $P_c^{IDLE}$ | Idle network power consumption in each IoT node $c$ in mW. |
| $E_{ef}^{PBT}$ | Energy per bit for each IoT link $(e, f)$ in mW/kbps. |
| M | Large number ($= 10^8$). |
| $P_e^{CAPT}$ | Link capacity for each IoT node $(e)$ in kbps. |
| $F_{ef}^{TR}$ | Transmit amplifier factor for each IoT link $(e, f)$ in mW/kbps/$m^2$. |
| Variables | |
| $I_{iac}^{NE}$ | $I_{iac}^{NE}$ is node embedding indicator, $I_{iac}^{NE} = 1$ If virtual node $a$ in BP $i$ has been embedded in IoT node $c$, $I_{iac}^{NE} = 0$ otherwise. |
| $I_{iacn}^{F}$ | $I_{iacn}^{F}$ is function embedding indicator, $I_{iacn}^{F} = 1$ if IoT node $c$ has the function $n$ required by virtual node $a$ in BP $i$, $I_{iacn}^{F} = 0$ otherwise. |
| $I_{iacz}^{Z}$ | $I_{iacz}^{Z}$ is zone embedding indicator, $I_{iacz}^{Z} = 1$ if IoT node $c$ is located in zone $z$ required by virtual node $a$ in BP $i$, $I_{iacz}^{Z} = 0$ otherwise. |
| $I_{iabcd}^{LE}$ | $I_{iabcd}^{LE}$ is link embedding indicator, $I_{iabcd}^{LE} = 1$ if the neighbouring virtual nodes $(a, b)$ in BP $i$ have been embedded in IoT nodes $(c, d)$, $I_{iabcd}^{LE} = 0$ otherwise. |
| $X_{iabcd}^{XOR}$ | Dummy binary variable |
| $R_{cd}^{TRFP}$ | Embedded traffic demand between IoT nodes $(c, d)$ in kbps. |
| $R1_{cde}^{TR}$ | Primary path between IoT nodes $(c, d)$ traversing the neighboring IoT nodes $(e, f)$ in kbps. |
| $R2_{cde}^{TR}$ | Secondary path between IoT nodes $(c, d)$ traversing the neighboring IoT nodes $(e, f)$ in kbps. |
| $I_{cdef}^{R1}$ | Primary path indicator, $I_{cdef}^{R1} = 1$ If the traffic demand between IoT nodes $(c, d)$ traverses neighboring IoT nodes$(e, f)$, $I_{cdef}^{R1} = 0$ otherwise. |
| $I_{cdef}^{R2}$ | Secondary path indicator, $I_{cdef}^{R2} = 1$ If the traffic demand between IoT nodes $(c, d)$ traverses neighboring IoT nodes$(e, f)$, $I_{cdef}^{R2} = 0$ otherwise. |
| $R_{ef}^{TRFL}$ | Traffic between neighboring IoT nodes $(e, f)$ in kbps. |
| $R_{ef}^{TRFL}$ | Traffic between neighboring IoT nodes $(e, f)$ in kbps. |
| $R_{f}^{TRFN}$ | Arrival rate of IoT nodes $(f)$ in kbps. |
| $LI_{fj}^{Lmb}$ | Lambda indicator for each IoT node $(f)$ with corresponding arrival rate $(j)$ then $LI_{fj}^{Lmbda} = 1$, otherwise 0. |
| $W_f^{NOD}$ | Traffic mean latency for each node $(f)$ in ms. |
| $I_c^{PM}$ | $I_c^{PM} = 1$ if the processing module indicator of IoT node $c$ is powered on, $P_c^{PM} = 0$ otherwise. |
| $I_c^{TM}$ | $I_c^{TM} = 1$ if the network module indicator of IoT node $c$ is powered on, $I_c^{TM} = 0$ otherwise. |
| $TPP$ | Total processing power consumption in the IoT network in mW. |
| $TNP$ | Total network power consumption in the IoT network in mW. |
| $TL$ | Total traffic mean latency in traffic the primary path in ms. |

## B. Framework objective function

The proposed framework minimizes the power consumption and the queuing latency in an IoT network by using the following objective function:

$$\textbf{Objective: minimize } \boldsymbol{\alpha.TL + \beta.TPP + \gamma.TNP} \qquad (1)$$

where α, β, and γ are the weight values thus used for magnitude and units. The framework selects the traffic value for each link in the network that preserves the low power consumption and the mean traffic latency at feasible values of the arrival rate. To enhance optimality of the power saving and latency minimization, we used the weight values given in our former work (α = 30/ms, β = 1/mW, and γ = 1/mW), [19].
Here, the total traffic latency for the IoT nodes can be calculated as follows:

$$TL = \sum_{f \in P} W_f^{NODE} \qquad (2)$$

where $W_f^{NODE}$ represents the average waiting time of the packets waiting to be processed for each IoT node in milliseconds according to queuing waiting time.
TPP is the total processing power and can be calculated as follows:

$$TPP = \sum_{c \in P} I_c^{PM} \cdot P_c^{IDLECP} \qquad (3)$$
$$+ \sum_{c \in P} \sum_{i \in B} \sum_{a \in V} I_{iac}^{NE} \cdot P_c^{MAXCP}$$
$$\cdot \frac{V_{ia}^{MCU}}{P_c^{MCU}}$$

where $I_c^{PM}$ is a binary variable that indicates an active processing module in IoT node c, $P_c^{IDLECP}$ is the idle processing power parameter of IoT node c in milliwatts, $I_{iac}^{NE}$ is a binary variable that indicates that virtual node a in BP $i$ has been embedded in IoT node $c$, $P_c^{MAXCP}$ is the parameter of maximum CPU power consumption in each IoT node $c$ in milliwatts, $V_{ia}^{MCU}$ is a parameter that specifies the processing requirement of virtual node $a$ in BP $a$ in megahertz, and $P_c^{MCU}$ is a parameter that specifies the processing capability of the IoT node $c$ in megahertz. The processing power consumption is considered to follow a linear profile versus the load with idle power consumption.
Here, the network power consumption in the IoT network can be expressed as follows:

$$TNP = \sum_{e \in P} I_e^{TM} \cdot P_e^{IDLETP} \qquad (4)$$
$$+ 2 \cdot \sum_{e \in PN} \sum_{f \in PB_e} R_{ef}^{TRFL1} \cdot E_{ef}^{PBT} + 2$$
$$\cdot \sum_{e \in PN} \sum_{f \in PB_e} R_{ef}^{TRFL2}$$
$$\cdot E_{ef}^{PBT}$$

$$+ \sum_{e \in PN} \sum_{f \in PB_e} R_{ef}^{TRFL1} \cdot (P_{ef}^{DIST})^2 \cdot F_{ef}^{TR}$$
$$+ \sum_{e \in PN} \sum_{f \in PB_e} R_{ef}^{TRFL2}$$
$$\cdot (P_{ef}^{DIST})^2 \cdot F_{ef}^{TR}$$

where $I_e^{TM}$ is a binary variable that indicates an active network module in IoT node $e$, $P_e^{IDLETP}$ is the idle network power parameter of IoT node $e$, $R_{ef}^{TRFL1}$ and $R_{ef}^{TRFL2}$ indicate the primary and alternative paths' traffic between neighboring IoT nodes $(e, f)$ in kb/s, $E_{ef}^{PBT}$ represents the energy per bit of each IoT link $(e, f)$ in milliwatts per kilobit per second, $P_{ef}^{DIST}$ denotes the distance between the neighboring IoT nodes pair $(e, f)$ in meters, and $F_{ef}^{TR}$ represents the transmit amplifier factor [45] for each IoT link $(e, f)$ in milliwatts per kilobit per second per metre square.

## C. Framework constraints

The proposed framework performs the embedding operation in two parts as follows:

### 1) Embedding of virtual nodes

$$\sum_{c \in P} I_{iac}^{NE} = 1 \tag{5}$$
$$\forall i \in B, \quad \forall a \in V$$
$$\sum_{a \in V} I_{iac}^{NE} \leq 1 \tag{6}$$
$$\forall i \in B, \forall c \in P$$

Constraint (5) ensures that each virtual node in a BP is embedded in a single IoT node only. Constraint (6) states that each IoT node is not allowed to host more than one virtual node in BP. This is considered the coexistence constraint and is not used in all the scenarios, such as controller node virtualization.

$$\sum_{i \in B} \sum_{a \in V} I_{iac}^{NE} \geq I_c^{PM} \tag{7}$$
$$\forall c \in P$$
$$\sum_{i \in B} \sum_{a \in V} I_{iac}^{NE} \leq I_c^{PM} \cdot M \tag{8}$$
$$\forall c \in P$$

Constraints (7) and (8) add a processing module in IoT node $c$ if this node is chosen for embedding at least one virtual node $a$ in BP $i$ or more, where $M$ is a sufficiently large unitless number to ensure that $P_c^{PMI} = 1$ when $\sum_{i \in B} \sum_{a \in V} P_{iac}^{NE}$ is greater than zero.

$$\sum_{i \in B} \sum_{a \in V} V_{ia}^{MCU} \cdot I_{iac}^{NE} \leq P_c^{MCU} \tag{9}$$
$$\forall c \in P$$
$$\sum_{i \in B} \sum_{a \in L} V_{ia}^{RAM} \cdot I_{iac}^{NE} \leq P_c^{RAM} \tag{10}$$
$$\forall c \in P$$

Constraints (9) and (10) represent the MCU and the memory capacity constraints, respectively. They ensure that the embedded MCU and memory workloads in an IoT node do not exceed the processor and memory capacities, respectively.

$$I_{iac}^{NE} \cdot V_{ian}^{FUNC} = I_{iacn}^{F} \tag{11}$$
$$P_{cn}^{FUNC} >= I_{iacn}^{F} \tag{12}$$
$$\forall i \in B, \forall a \in L, \forall c \in P, \forall n \in F$$

Constraints (11) and (12) ensure that the required function of each virtual node in a BP is provided by its hosting IoT node.

$$I_{iac}^{NE} \cdot V_{iaz}^{ZONE} = I_{iacz}^{Z} \tag{13}$$
$$P_{cz}^{ZONE} \geq I_{iacz}^{Z} \tag{14}$$
$$\forall i \in B, \forall a \in V, \forall c \in P, \forall z \in Z$$

Constraints (13) and (14) ensure that the required zone of each virtual node in BP is matched by the zone of the hosting IoT node.

### 2) Embedding of virtual links

$$I_{iac}^{NE} + I_{ibd}^{NE} = X_{iabcd}^{LE} + 2 \cdot I_{iabcd}^{LE} \tag{15}$$
$$\forall i \in B, \forall a \in V, \forall b \in VN_{ia} : a \neq b, \forall c, d \in P : c \neq d$$

Constraint (15) ensures that neighboring virtual nodes $a$ and $b$ of $i$ in $B$ are also connected in embedding IoT nodes $c$ and $d$. We achieved this by introducing a binary variable $I_{iabcd}^{LE}$, which is only equal to 1 if $I_{iac}^{NE}$ and $I_{ibd}^{NE}$ are exclusively equal to 1; otherwise, it is zero, when $X_{iabcd}^{LE}$ is a neglected variable.

$$\sum_{i \in B} \sum_{a \in L} \sum_{b \in LNB_{ia}} I_{iabcd}^{LE} \cdot V_{iab}^{TRFIC} = R_{cd}^{TRFP} \tag{16}$$
$$c, d \in P : c \neq d$$

Constraint (16) generates the path's traffic matrix resulting from embedding virtual nodes $a$ and $b$ into IoT nodes $c$ and $d$.

#### a) Retransmission- and replication-based schemes

In this scheme, the proposed framework finds two energy-efficient routes for the traffic between the embedded nodes, namely the primary and alternative routes.

$$\sum_{f \in PN_e} R_{cdef}^{TR1} \tag{17}$$
$$- \sum_{f \in PN_e} R_{cdfe}^{TR1} \begin{cases} R_{cd}^{TRFP} & if\, e = c \\ -R_{cd}^{TRFP} & if\, e = d \\ 0 & otherwise \end{cases}$$
$$\forall c, d, e \in P : c \neq d \text{ and } e \neq f$$

Constraint (17) represents the flow conservation constraint for the traffic flows in the IoT network.

$$\sum_{c \in P} \sum_{d \in P} R_{cdef}^{TR1} = R_{ef}^{TRFL1} \tag{18}$$
$$\forall e \in P, \forall f \in PN_e$$

Constraint (18) generates a link's traffic matrix between the neighboring IoT nodes $e$ and $f$.

$$R_{cdef}^{TR1} \geq I_{cdef}^{R1} \tag{19}$$

$$R_{cdef}^{TR1} \leq I_{cdef}^{R1} \cdot M \tag{20}$$

$$\forall c, d, e \in PN, \forall f \in PB_e : c \neq d, e \neq f$$

Constraints (19) and (20) build the primary path indicator between embedding IoT nodes $c$ and $d$ through neighboring IoT nodes $e$ and $f$, where $I_{cdef}^{R1} = 1$ if there is a traffic path between IoT nodes $c$ and $d$ that passes through neighboring IoT nodes $e$ and $f$, where M is a sufficiently large unitless number to ensure that $R_{cdef}^{R1} = 1$ when $R_{cdef}^{ROUTE1}$ is greater than zero.

$$\sum_{f \in PB_e} I_{cdef}^{R1} \leq 1 \tag{21}$$

$$\forall c, d, e \in PN : c \neq d \text{ and } e \neq f$$

Constraint (21) ensures that traffic splitting is prevented for each path between embedding IoT nodes $c$ and $d$, such that the maximum number of physical links between neighboring IoT nodes $e$ and $f$ is one.

$$\sum_{f \in PN_e} R_{cdef}^{TR2}$$

$$- \sum_{f \in PN_e} R_{cdfe}^{TR2} \begin{cases} R_{cd}^{TRFP} & \text{if } e = c \\ -R_{cd}^{TRFP} & \text{if } e = d \\ 0 & \text{otherwise} \end{cases} \tag{22}$$

$$\forall c, d, e \in PN : c \neq d \text{ and } e \neq f$$

Constraint (22) represents the flow conservation constraint for the alternative path's traffic flows in the IoT network.

$$\sum_{c \in P} \sum_{d \in P} R_{cdef}^{TR2} = R_{ef}^{TRFL2} \tag{23}$$

$$\forall e \in PN, \forall f \in PB_e$$

Constraint (23) generates the alternative link's traffic matrix between neighboring IoT nodes $e$ and $f$.

$$R_{cdef}^{TR2} \geq I_{cdef}^{R2} \tag{24}$$

$$R_{cdef}^{TR2} \leq I_{cdef}^{R2} \cdot M \tag{25}$$

$$\forall c, d, e \in PN, \forall f \in PB_e : c \neq d, e \neq f$$

Constraints (24) and (25) build the alternative path between embedding IoT nodes $c$ and $d$ through neighboring IoT nodes $e$ and $f$, where $R_{cdef}^{R2} = 1$ if there is a traffic path between IoT nodes $c$ and $d$ that passes through neighboring IoT nodes $e$ and $f$, where $M$ is a sufficiently large unitless number to ensure that $I_{cdef}^{TR2} = 1$ when $R_{cdef}^{TR2}$ is greater than zero.

$$\sum_{f \in PB_e} R_{cdef}^{R2} \leq 1 \tag{26}$$

$$\forall c, d, e \in PN : c \neq d \text{ and } e \neq f$$

Constraint (26) ensures that traffic splitting is prevented for each path between embedding IoT nodes $c$ and $d$, such that the maximum number of physical links between neighboring IoT nodes $e$ and $f$ is one.

$$I_{cdef}^{R1} + I_{cdef}^{R2} \leq 1 \tag{27}$$

$$\forall c, d, e \in PN, \forall f \in PB_e : c \neq d, e \neq f$$

Constraint (27) ensures the traffic creation of two distinct paths between embedding IoT nodes $c$ and $d$ such that each path uses different physical links between neighboring IoT nodes $e$ and $f$.

$$\sum_{c \in PN} \sum_{d \in PN} \sum_{f \in PB_e} I_{cdef}^{R1} + I_{cdef}^{R2} \geq I_e^{TM} \tag{28}$$

$$\sum_{c \in PN} \sum_{d \in PN} \sum_{f \in PB_e} R_{cdef}^{R1} + R_{cdef}^{R2} \leq I_e^{TM} \cdot M \tag{29}$$

$$e \in PN : c \neq d \text{ and } e \neq f$$

Constraints (28) and (29) build a network module indicator of IoT node $e$ if this IoT node is chosen for send/receive traffic for at least one link or more, where $M$ is a sufficiently large unitless number to ensure that $I_e^{TM} = 1$ when $\sum_{c \in PN} \sum_{d \in PN} \sum_{f \in PB_e} I_{cdef}^{R1} + I_{cdef}^{R2}$ is greater than zero.

$$\sum_{e \in PN_f} R_{ef}^{TRFL1} + R_{ef}^{TRFL2} = R_f^{TRFN} \tag{30}$$

$$\forall f \in P : e \neq f$$

Constraint (30) estimates the arrival traffic for each IoT node.

$$\sum_{f \in P} R_f^{TRFN} \leq CAPACITY \tag{31}$$

Constraint (31) states that the total traffic flow of the IoT node $f$ should not exceed the node capacity.

$$\sum_{j \in J} LI_{f\,j}^{LMBDA} \cdot j = R_f^{TRFN} \tag{32}$$

$$\forall f \in P$$

Constraint (32) determines the arrival rate for each IoT node.

$$\sum_{j \in J} LI_{f\,j}^{LMBDA} \leq 1 \tag{33}$$

$$\forall f \in P$$

Constraint (33) ensures that each IoT node has no more than one arrival rate indicator.

$$\sum_{j \in J} W_j^{LIMDA} \cdot LI_{f\,j}^{LMBDA} = W_f^{NODE} \tag{34}$$

$$\forall f \in P$$

Constraint (34) estimates the traffic delay for each IoT node $f$ on the basis of the product of the lambda indicator and the corresponding latency for this lambda $j$.

### b) Splitting-based schemes

In this section, we propose a traffic splitting-based resilience scheme through the multiple paths concept to reduce the arrival rates through the intermediate nodes; doing so will consequently minimize the delivery time, in addition to enhancing the resilience of the IoT network.

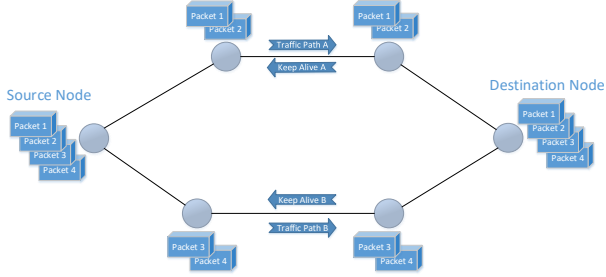

Fig 2: Traffic Splitting Scheme

The proposed framework splits the traffic between the source node and the destination node and routes it into two paths (A and B), as shown in Fig. 2. The source node sends one half of the traffic through path A and the other half through path B to the destination node, and the source node receives a 'Keep-alive' signal continuously from both paths (A and B). Once a failure occurs on one path, the source will not receive an acknowledgement from this path and will then switch the traffic to another path.

Let us suppose that the source node has 100 packets to send to the destination node. The source node selects two paths and sends 50 packets on each path to the destination node. In a probabilistic scenario in which one link has failed, the source node will resend only 50 packets or less rather than resending all 100 packets as in retransmission.

In this scheme, the proposed framework finds the two best routes in terms of energy-efficiency for the traffic between the embedded nodes, namely the primary and the secondary routes. The main difference between this splitting scheme and the former schemes is the flow conservation constraints in (17) and (22).

$$\sum_{f \in PN_e} P_{cdef}^{ROUTE1} \quad (35)$$

$$- \sum_{f \in PN_e} P_{cdfe}^{ROUTE1} \begin{cases} 0.5 \cdot P_{cd}^{TRFP} & \text{if } e = c \\ -0.5 \cdot P_{cd}^{TRFP} & \text{if } e = d \\ 0 & \text{otherwise} \end{cases}$$

$$\forall c, d, e \in PN : c \neq d \text{ and } e \neq f$$

$$\sum_{f \in PN_e} P_{cdef}^{ROUTE2} \quad (36)$$

$$- \sum_{f \in PN_e} P_{cdfe}^{ROUTE2} \begin{cases} 0.5 \cdot P_{cd}^{TRFP} & \text{if } e = c \\ -0.5 \cdot P_{cd}^{TRFP} & \text{if } e = d \\ 0 & \text{otherwise} \end{cases}$$

$$\forall c, d, e \in PN : c \neq d \text{ and } e \neq f$$

Constraints (35) and (36) represent the flow conservation constraints for the primary and secondary paths for the traffic splitting scheme.

## IV. RESULTS AND DISCUSSION

To evaluate the performance of the proposed model, we considered a smart building scheme (i.e. enterprise or university campus) where the physical layer is composed of 30 IoT nodes connected by 89 bidirectional wireless links. These IoT nodes are distributed randomly in buildings across a campus in an area of 500 m × 500 m as shown in Fig. 3.

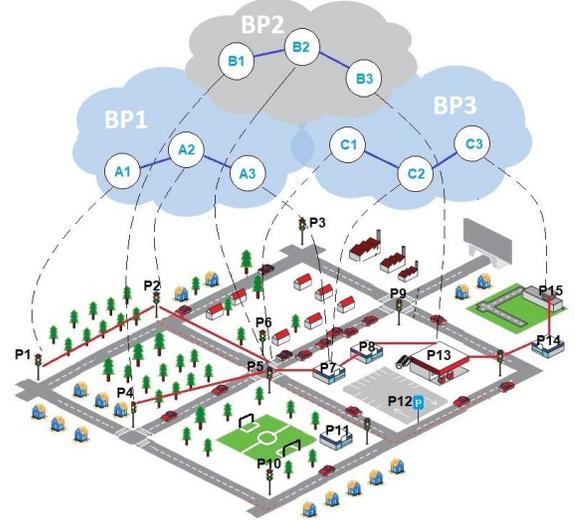

Fig 3: Service embedding layers in IoT networks

We evaluated the power consumption and the mean traffic latency resulting from resilient service embedding across distinct zones with the coexistence constraint. The model considered the objective function discussed in Section III–B for energy efficient-low latency service embedding. Table 2 and 3 list the model input parameters [46]. A comprehensive description of the setup and the processors used in each IoT node can be found in our previous work in [19].

Table 1: MILP model input parameters

| Parameter Description | Value and Unit |
|---|---|
| Energy per bit | 50 nJ/bit |
| Maximum traffic capacity of node | 250 kb/s |
| Packet size | 128 byte |
| Maximum link distance | 100 m |
| Transmitter amplifier power coefficient | 255 pJ/bit. $m^2$ |
| Scale factor with large value (M) | 1000000 |

Table 2: Processing modules power specifications and power consumption in active mode

| MCU Type | MCU CLK | Idle Power | Max. Power |
|---|---|---|---|
| MSP430F1 | 8 MHz | 1 mW | 8 mW |
| MSP430FR5 | 16 MHz | 1 mW | 14 mW |
| MSP430FR6 | 16 MHz | 1 mW | 20 mW |
| MSP430F5 | 25 MHz | 1 mW | 14 mW |
| MSP432P4 | 48 MHz | 1 mW | 16 mW |

The probabilistic model is based on k-connected nodes with the assumption that the network has the ability to recover from failures in the case of a link or node failure. We use our model to evaluate two resilience schemes:

## A. Energy-efficient low-latency node-resilient service embedding

For the node-resilient scheme, we run three resilience levels with the objective of minimizing the total power consumption and the mean traffic latency:

- Coexistence constraint node resilience (CCNR)
- Partial redundancy node resilience (PRNR)
- Full redundancy node resilience (FRNR)

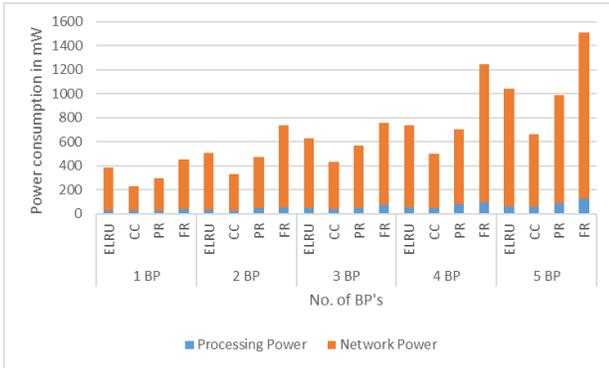

Fig 4: Power consumption of energy-efficient low-latency node-resilient service embedding.

The results shown in Fig. 4 show the total power consumption of CCNR, PRNR, and FRNR and compares them with the energy-latency-resilience unaware (ELRU) scenario. These results demonstrate that the CCNR scenario has an average power saving of 35% compared with the ELRU scenario. While the higher level of power consumption in the PRNR scenario has an average power saving of 10% compared with ELRU.

The FRNR has higher power consumption than the other scenarios, and the average power consumption is 40% higher than that in the ELRU scenario.

The increase in power consumption in each scenario is due to the embedding of the redundant nodes and the traffic among these nodes, but the node resilience level is improved and the IoT network has the ability to maintain service provisioning even with a failure in one node.

## B. Energy-efficient low-latency traffic-resilient service embedding

For the traffic-resilient scheme, we run three resilience levels with the objective of minimizing the total power consumption and the traffic mean latency:

- Redundancy-based traffic resilience (RDTR)
- Replication-based traffic resilience (RPTR)
- Splitting-based traffic resilience (STR)

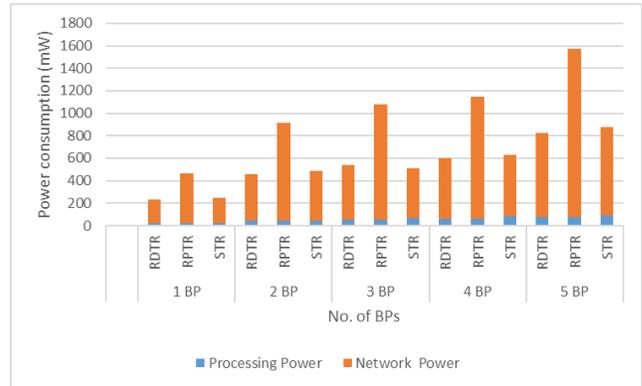

Fig 5: Power consumption of traffic-resilient service embedding scenarios without failure.

The results presented in Fig. 5 display the power consumption of the traffic-resilient service embedding for the RDTR, RPTR, and STR scenarios in the packet delivery case without a failure. These results show that RDTR has the lowest power consumption with an average power saving of 47% and 4% compared with RPTR and STR scenarios, respectively. Notice than in some cases (i.e. 3 BP's embedding), the STR has lower power consumption compared with RDTR. This is due to its ability to find energy efficient routes for part of the traffic, i.e. 50 % of the total traffic.

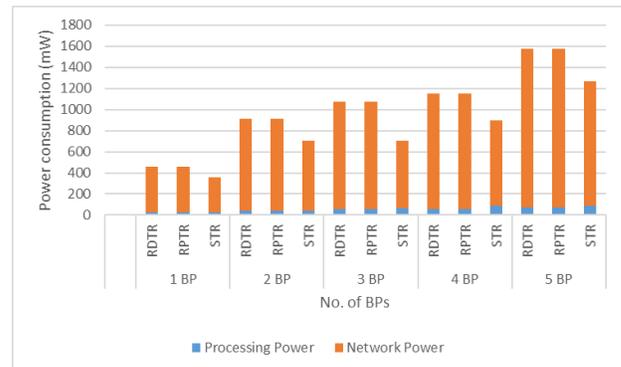

Fig 6: Power consumption of traffic-resilient service embedding scenarios with failure.

The results presented in Fig. 6 show the power consumption of the traffic-resilient service embedding for the RDTR, RPTR, and STR scenarios in the packet delivery case with one link failure. These results reveal that RDTR has the same power consumption as RPTR because of the data retransmission through the secondary path. The results also reveal that the STR has an average power saving of 25% compared with the RDTR scenario. These results show that the proposed technique in the STR scenario has higher power consumption by 4%, but 25% power savings in the case of one link failure.

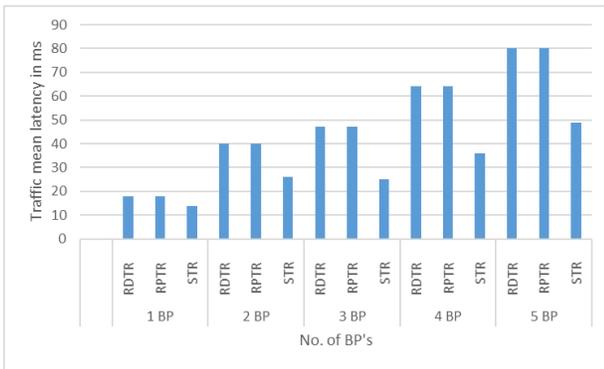

Fig 7: Traffic mean latency of traffic resilient-service embedding scenarios without failure.

The results presented in Fig. 7 show the mean network traffic latency of the service embedding scenarios. These results demonstrate that the STR reduces the average mean traffic latency by 37% for the set of parameters used, compared with the RDTR and RPTR scenarios. The mean traffic latency minimization in STR is due to the traffic splitting and hence the reduction in the arrival rate of the individual nodes. The traffic splitting technique offered better performance in terms of the end-to-end delay.

The packet delivery ratio (PDR) reflects the network performance level, where better network performance resulted in a high packet delivery ratio. The packet delivery ratio is inversely proportional to the network size in IoT networks because the routing performance is better in a low-node-density network.

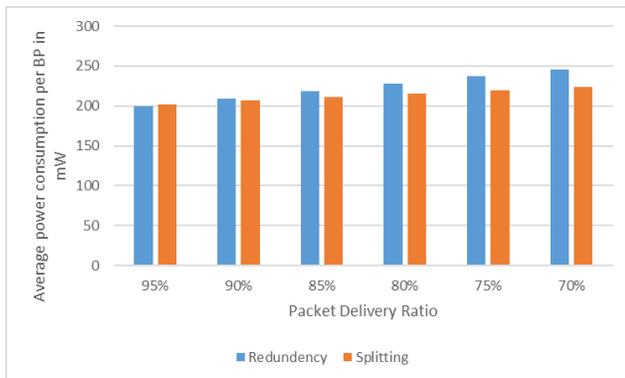

Fig 8: Power consumption of traffic-resilient service embedding scenarios for different PDR scenarios.

The results shown in Fig.8 present a comparison of the total power consumption in the RDTR and STR scenarios for different PDR values [47]. These results demonstrate that the RDTR is an energy-efficient technique for high-performance networks (i.e. PDR > 95%). However, the STR scenario produces higher power savings with lower PDR. The STR scenario exhibits power savings of 10% compared with RDTR when PDR = 70%. These results help in comparing the RDTR and the STR without the RPTR, where the RPTR has the highest power consumption in all the cases.

## V. SUMMARY

In this paper, multilevel node and traffic resilience schemes for IoT networks were reviewed. A MILP model was developed to enhance the resilience of the services offered. The node and traffic resilience were enhanced by using the proposed scheme and a model for energy-efficient low-latency resilient service embedded in a smart building was developed. A range of node and traffic resilience levels were developed and their performance in terms of the mean traffic latency and power consumption were compared. A novel technique was also proposed based on traffic splitting to enhance the network resilience and performance by reducing the packet delivery time. Moreover, splitting techniques were evaluated using redundancy and replication resilience techniques in terms of the total power consumption and the mean traffic latency for different values of PDR.

The results showed that the STR scenario produced higher power savings with lower PDR. The STR scenario exhibited a power saving of 10% compared with the RDTR scheme when PDR was equal 70%. The results also revealed that the STR reduced the average mean traffic latency by 37% compared with the RDTR and RPTR scenarios. The mean traffic latency minimization in STR was due to traffic splitting which reduced the traffic arrival rate at the nodes. The traffic splitting technique also exhibited better performance in terms of the end-to-end delay.

## VI. ACKNOWLEDGEMENTS

The authors would like to acknowledge funding from the Engineering and Physical Sciences Research Council (EPSRC), INTERNET (EP/H040536/1), STAR (EP/K016873/1) and TOWS (EP/S016570/1) projects. Mr. Haider Al-Shammari would like to thank the Higher Committee for Education Development (HCED) for funding his scholarship.